\documentclass{jfm}

\usepackage{graphicx}
\usepackage{amssymb}
\usepackage{amsfonts}
\usepackage{natbib}
\usepackage{pstool}
\usepackage{xcolor}
\usepackage{subfigure}
\usepackage{verbatim}
\usepackage{pifont}
\usepackage{psfrag}

\title[Scaling of turbulence in pipe flow]{Scaling turbulence in the near-wall region}

\author[Hultmark and Smits]%
{M\ls A\ls R\ls C\ls U\ls S\ns H\ls U\ls L\ls T\ls M\ls A\ls R\ls K  \and A\ls
L\ls E\ls X\ls A\ls N\ls D\ls E\ls R\ns J.\ns S\ls M\ls I\ls T\ls S}

\affiliation{Department of Mechanical and Aerospace Engineering, 
Princeton University, \break
Princeton, NJ 08544, USA\\[\affilskip]
}

\pubyear{2021}
\volume{}
\pagerange{}

\begin{document}

\maketitle

\begin{abstract}
A new velocity scale is derived that yields a Reynolds number independent profile for the streamwise turbulent fluctuations in the near-wall region of wall bounded flows for $y^+<25$.  The scaling demonstrates the important role played by the wall shear stress fluctuations and how the large eddies determine the Reynolds number dependence of the near-wall turbulence distribution. 

\end{abstract}

\section{Introduction}
\label{intro}

For isothermal, incompressible flow, it is commonly assumed that for the region close to the wall
$$[U, \overline{u^2}] = f(y, u_\tau, \nu, \delta),$$
where $U$ and $u$ are the mean and fluctuating velocities in the streamwise direction $x$.  The overbar denotes ensemble averaging, and the outer length scale $\delta$ is, as appropriate, the boundary layer thickness, the pipe radius, or the channel half-height.  That is,
\begin{equation}
[{U^+},\overline{{u^2}^+}] = f(y^+, Re_\tau).
\label{dim_analysis}
\end{equation}
where the friction Reynolds number $Re_\tau=\delta u_\tau/\nu$, and $\overline{{u^2}^+}=\overline{{u^2}}/u_\tau^2$.  Here, $y^+=yu_\tau/\nu$, and the superscript $^+$ denotes non-dimensionalization using the fluid kinematic viscosity  $\nu$ and the friction velocity $u_\tau=\sqrt{\tau_w/\rho}$, where $\tau_w$ is the wall shear stress and $\rho$ is the fluid density. 

\begin{figure}
\centering
\includegraphics[width=0.55\textwidth]{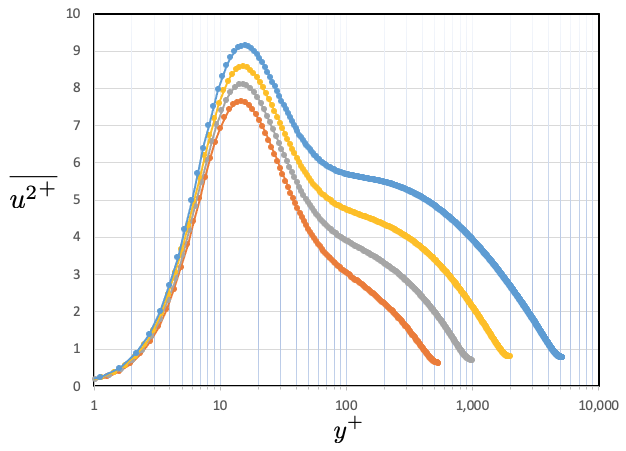} 
\caption{ Profiles of $\overline{{u^2}^+}$ for $Re_\tau =$ 550, 1000, 2000, 5200. Channel flow DNS \citep{lee2015}.  }
\label{u2}
\end{figure}

By all experimental indications, the mean flow is a unique function of $y^+$ in the inner region that is independent of Reynolds number for $y/\delta \lessapprox 0.15$ (see, for example, Zagarola \& Smits 1998, and McKeon {\it et al.\/} 2004). \nocite{Zagarola1998, McKeon:2004}  In contrast, the Reynolds stresses have been shown to exhibit a dependence on Reynolds number in the near-wall region, as illustrated in figure~\ref{u2} for $\overline{{u^2}^+}$. In particular, the peak turbulence intensity at $y^+ \approx 15$ increases with Reynolds number, and \cite{samie2018fully} showed that over a wide range of Reynolds numbers the magnitude of the peak in a boundary layer follows a logarithmic variation
\begin{equation}
\overline{{u^+_p}^2} = 3.54 + 0.646 \ln(Re_\tau).
\label{up1}
\end{equation}
\cite{lee2015} found a very similar result from DNS of a channel flow (using only the data for $y^+ \ge 1000$)
\begin{equation}
\overline{{u^+_p}^2} = 3.66 + 0.642 \ln(Re_\tau),
\label{up2}
\end{equation}
very much in line with the result reported by \cite{Lozano2014}, also obtained by DNS of channel flow, where
$$\overline{{u^+_p}^2} = 3.63 + 0.65 \ln(Re_\tau).$$
\cite{chen2021reynolds} have argued that these logarithmic increases cannot be sustained because the dissipation rate at the wall is bounded. Nevertheless, a non-trivial Reynolds number dependence is indicated by all accounts for finite Reynolds numbers.

Here, the aim is to better understand this Reynolds number dependence. A new velocity scale is proposed that yields a universal profile for the fluctuations in the region $y^+<25$, which includes the inner peak in the near-wall stress profile.  The analysis draws primarily on the direct numerical simulation (DNS) results for channel flow reported by \cite{lee2015}, but it is expected to hold more generally for two-dimensional zero-pressure gradient boundary layers and fully developed pipe flows. 

\section{Taylor Series expansion}
We start by writing the Taylor series expansions for the streamwise velocity in the vicinity of the wall \citep{bewley2004skin}, where
\begin{equation}
\underline u^+  =  a_1 + b_1 y^+  + c_1{y^+}^2 + O({y^+}^3). \label{Taylor1u} 
\end{equation}
The underline indicates an instantaneous value such that $\underline u=U+u$.  The no-slip condition gives $a_1=0$, and
\begin{eqnarray}
b_1 & = &  (\partial \underline  u^+/ \partial y^+)_w  \label{b1}  \\
c_1 & = & {\textstyle {1 \over 2}} (\partial \underline p^+/ \partial x^+)   \label{c1}
\end{eqnarray}

\begin{figure}
\centering
\includegraphics[width=0.47\textwidth]{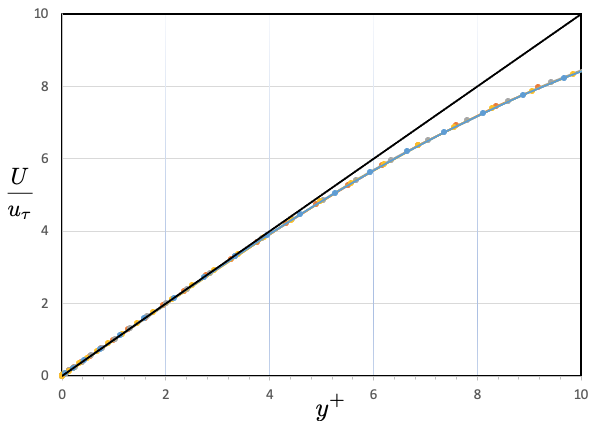}
\includegraphics[width=0.48\textwidth]{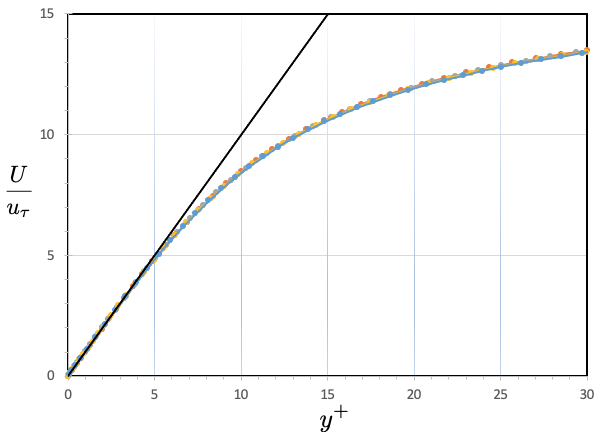}
\caption{ Mean velocity profiles.  $Re_\tau =$ 550, 1000, 2000, 5200 \citep{lee2015}.  }
\label{U_DNS}
\end{figure}

For the mean flow very close to the wall we recover the linear velocity profile
\begin{equation}
U^+=y^+.
\label{linear1}
\end{equation}
The channel flow results given in figure~\ref{U_DNS} shows that the linear profile is a good fit to the data for $y^+ < 4$, and as expected the data collapse very well for $y^+<100$.  

\begin{figure}
\centering
\includegraphics[width=0.47\textwidth]{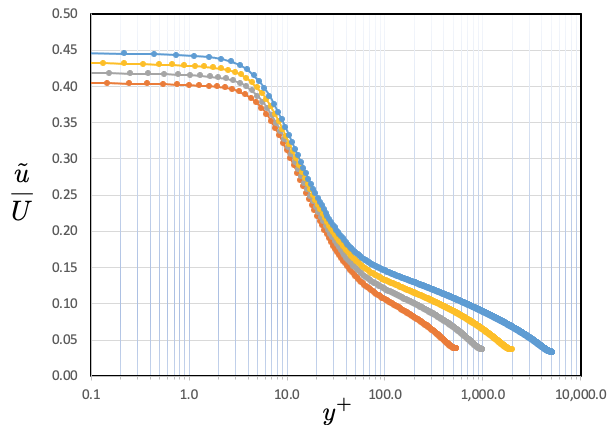}
\includegraphics[width=0.49\textwidth]{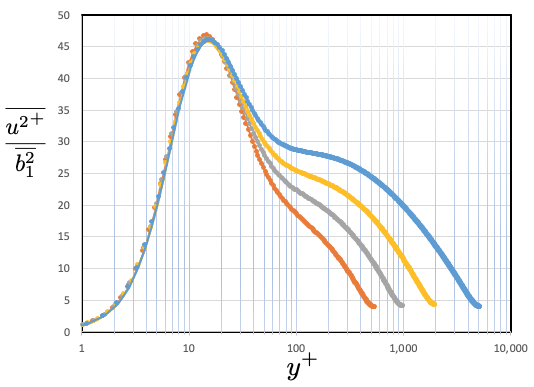}
\caption{ Left: Profiles of $\tilde u/U$; intercept at $y^+=0$ is $\tilde b_{1}$. Right: Profiles of $\overline{{u^2}^+}$ scaled with $\overline{b_1^2}$. $Re_\tau =$ 550, 1000, 2000, 5200 \citep{lee2015}.  }
\label{u_U_DNS}
\end{figure}

For the time-averaged fluctuations
\begin{equation}
\overline{{u^2}^+} = \overline{{b_1}^2}{y^+}^2 + \dots \label{Taylor1u2}
\end{equation}
This expansion only holds very close to the wall.  Since the mean velocity is linear in this region 
\begin{equation}
\tilde u/U  =  \tilde b_{1} \label{Taylor_rms_u}, 
\label{Taylor_rms_uv}
\end{equation}
where $ \tilde u = \sqrt{\overline{{u^2}^+}}$ and $ \tilde b_1 = \sqrt{\overline{b_1^2}}$.  The profiles of $\tilde u/U$ are displayed in figure~\ref{u_U_DNS} (left), and for the region $y^+<1$ we see that $b_1$ takes on a constant value. These values are listed in table~\ref{constants} for each Reynolds number.

\section{Magnitude of the inner peak}

If we now scale each $\overline{{u^+}^2} $ profile with the value of $\overline{b_1^2}$ at the same Reynolds number, we obtain the results shown in figure~\ref{u_U_DNS} (right).  The collapse of the data for $y^+<20$ is impressive, including the almost perfect agreement on the scaled inner peak value.  In fact, from table~\ref{constants} it is evident that $\overline{{b_1}^2}$ and $\overline{{u_p^+}^2}$ approach a constant ratio to each other such that
\begin{equation}
\overline{{u_p^+}^2} \approx 46\overline{{b_1}^2}.
\label{u2b12}
\end{equation} 
(using equation~\ref{up2}).  In other words, the magnitude of the peak at $y^+ \approx 15$ tracks almost precisely with $\overline{{b_1}^2}$, a quantity that is evaluated at $y^+=0$.

\begin{table}
\centering
\begin{tabular}{cccccc}
    \hspace{2mm} $Re_\tau$ \hspace{2mm} & \hspace{2mm} $\tilde b_1$ \hspace{2mm} & \hspace{2mm}   $\overline{{b_1}^2}$  \hspace{2mm} & \hspace{2mm}    $\overline{{u^+_p}^2}$  \hspace{2mm}  &  \hspace{2mm} $\overline{{u^+_p}^2} /\overline{{b_1}^2}$  \hspace{2mm}   & \hspace{2mm}  $b_1^2 u_\tau/U_{av} \times 10^3$  \hspace{2mm}    \\[1mm]
     550 	& 0.4046	&  0.1637  &  7.65 &  47.10 & 8.90 \\
     1000	& 0.4187	&  0.1753 &	8.10  &  46.17 & 8.77 \\
     2000	& 0.4324	&  0.1870  &	8.58 &  45.67 & 8.58 \\
     5200	& 0.4460	&  0.1989 &	9.14 &  46.01 & 8.25 
\end{tabular} 
\caption{Ratio $\overline{{u^+_p}^2} /\overline{{b_1}^2}$. Channel flow DNS \citep{lee2015}. }
\label{constants}
\end{table} 

We can now make some observations on mixed flow scaling.  \cite{DeGraaff2000} proposed that the correct scaling for $\overline{u^2}$ in boundary layers should be $\overline{{u^2}^+}\sqrt{C_f/2}$, where the skin friction coefficient $C_f= {\textstyle 1 \over 2}\tau_w/\rho U_e^2$ and $U_e$ is the mean velocity at the edge of the layer.  This is equivalent to scaling $\overline{u^2}$ with $u_\tau U_e$ (hence the term mixed scaling).  If this is correct, then $b_1^2 u_\tau/U_e$ needs to be invariant with Reynolds number.  As seen from table~\ref{constants}, this is not so, and despite the apparent collapse of their data for $540 < Re_\tau <$ 10,000 using $u_\tau U_e$, mixed scaling is only an approximation to the correct scaling. 

\section{What does it all mean?}

\begin{figure}
\centering
\includegraphics[width=0.55\textwidth]{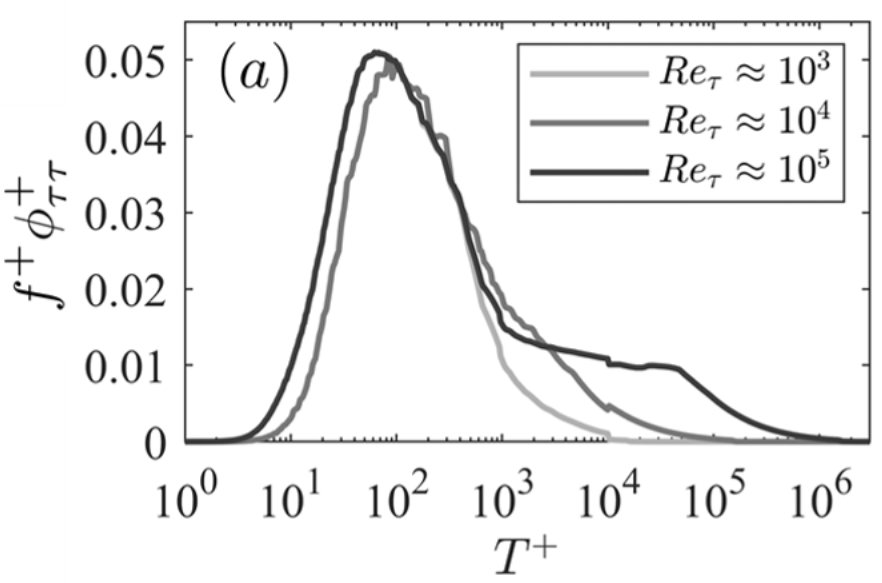} 
\caption{ Pre-multiplied spectra of the wall stress fluctuations $\tau'_w$ at $Re_\tau \approx$ 10$^3$, 10$^4$, and 10$^5$ ($T^+=1/f^+=u_\tau^2/(f \nu)$, $f$ is the frequency in Hz). Data taken from DNS of \cite{Del'Alamo2004}, hot-wire measurements, and predictive models \citep{Marusic2010a, mathis2013estimating, chandran2020spectral}. From \cite{Marusic2021submitted}. }
\label{wall_stress_spectra}
\end{figure}

Consider the constant $b_1$.  By equation~\ref{b1}, 
$$b_1 =  \left. \frac{\partial  \underline u^+}{\partial y^+} \right|_w = \frac{\nu}{u_\tau^2} \left. \frac{\partial  \underline u}{\partial y} \right|_w, $$
so that 
\begin{equation}
\overline{b_1^2} = \frac{1}{\rho^2 u_\tau^4} \overline{\left( \mu \frac{\partial u}{\partial y} \right)^2_w} = \frac{\overline{{\tau'_{w}}^2}}{\tau_w^2}.  
\label{uadef}
\end{equation}
The controlling parameter in the near-wall scaling for $\overline{{u_p^+}^2}$ is therefore the mean square of the fluctuating wall stress $\tau'_{w}$ in the $x$-direction.   
This scaling with $\tau'_{w}$ helps explain its Reynolds number dependence, since with increasing Reynolds number the large-scale (outer layer) motions contribute more and more to the fluctuating wall stress by modulation and superimposition of the near-wall motions \citep{Marusic2010a, agostini2018impact}, as may be seen from the pre-multiplied wall stress spectra shown in figure~\ref{wall_stress_spectra}.  Of course, it is the turbulence that controls the wall stress, and not vice versa, but the main point is that the whole of the region $y^+ < 20 $ (including the peak in $\overline{{u^2}^+}$) scales with the velocity scale $u_s=\tilde b_1 u_\tau$, which can be determined by measuring the fluctuating wall stress, a clear indication of the increasingly important contribution of the large-scale motions on the near wall behavior as the Reynolds number increases.   

\cite{chen2021reynolds} used a scaling argument to arrive at a result that is similar to that given in equation~\ref{u2b12}, and they related $b_1$ to the dissipation rate at the wall. However, $b_1$ only represents one part of the full dissipation rate $\mu \overline{\left({{\partial u_j} / {\partial x_i}} \right)^2}$, and we prefer to interpret $b_1$ in terms of streamwise component of the wall stress fluctuations. If the flow was isotropic, the connection between these two scaling arguments would be trivial, but the strongly anisotropic nature of the turbulence very close to the wall makes them fundamentally different.

\begin{figure}
\centering
\includegraphics[width=0.48\textwidth]{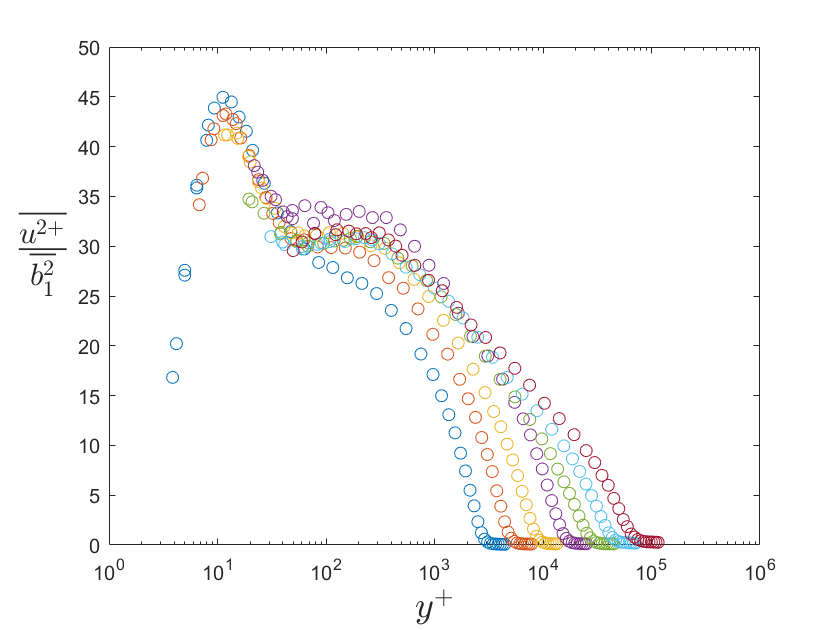} 
\includegraphics[width=0.48\textwidth]{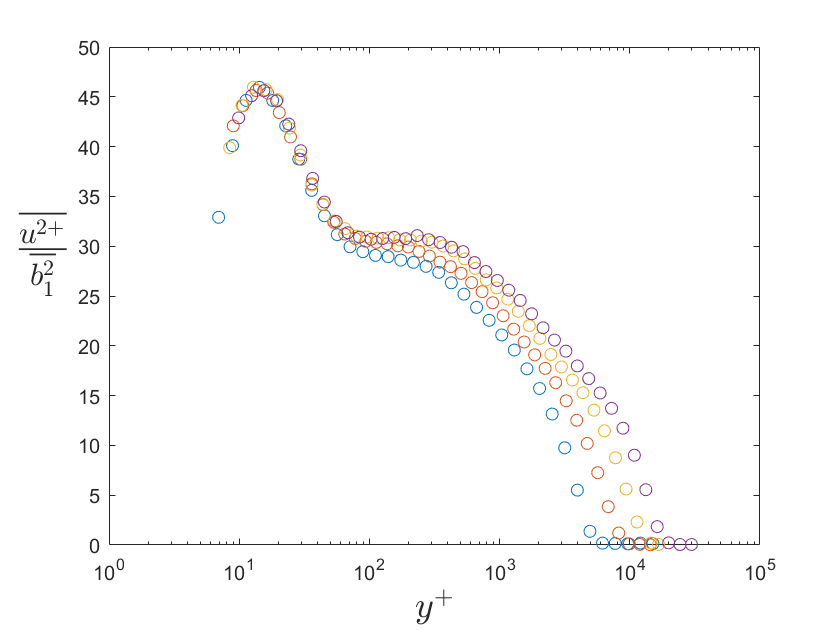} 
\caption{Scaled boundary layer experiments.   Left: From \cite{Vallikivi2015_HRTF}  for $Re_\tau =$ 2,622 to 72,526. Right: From \cite{samie2018fully} for $Re_\tau =$ 6,123 to 19,680. }
\label{HiRe_data}
\end{figure}

The analysis presented here relied on DNS of channel flow, which gives data for $y^+<1$ so that $\tilde b_1$ can be measured directly.  Such spatial resolution cannot be achieved in boundary layer measurements except at very low Reynolds numbers (see, for example, DeGraafff \& Eaton 2000), and so it is not usually possible to find $\tilde b_1$ by experiment.  \nocite{DeGraaff2000} If we assume that equation~\ref{u2b12} continues to hold at all Reynolds numbers, however, then  together with equation~\ref{up1} the value of $\tilde b_1$ can be estimated at each Reynolds number.  Figure~\ref{HiRe_data} shows the results for two sets of high Reynolds number boundary layer data, and it would appear that the data support the scaling suggested here on the basis of channel flow.  

\section{Back to $u/U$}

\begin{figure}
\centering
\includegraphics[width=0.49\textwidth]{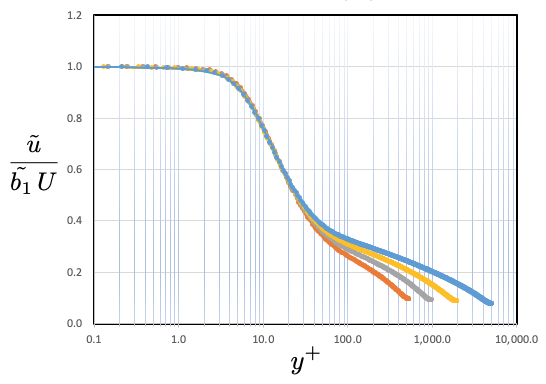}
\includegraphics[width=0.49\textwidth]{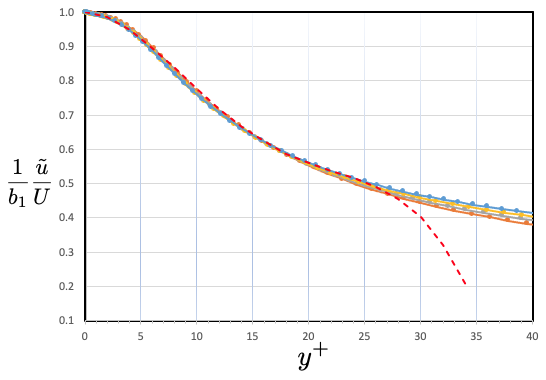}
\caption{Left: Profiles of $\tilde u/(\tilde b_1 U)$. Right: Profiles of $\tilde u/(\tilde b_1 U)$ showing equation~\ref{extendedTS} with $c'=-4.1 \times 10^{-3}$, $d'=2.21 \times 10^{-4}$ and $e'=-3.55 \times 10^{-6}$.  $Re_\tau =$ 550, 1000, 2000, 5200 \citep{lee2015}.  }
\label{logudivU}
\end{figure}

Figure~\ref{logudivU} shows the $\tilde u/U$ profiles from figure~\ref{u_U_DNS} (left) in log-log scaling.  All the data collapse onto a universal curve for $0<y^+<25$ (see also figure~\ref{u_U_DNS} (right)).  That is, the similarity revealed by the $u_s$ scaling extends over the whole viscous sublayer.

The form of this universal curve could be represented by an extension of the Taylor series given in equation~\ref{Taylor_rms_u} to higher order as in, for example,   
\begin{equation}
\tilde u/U  =  \tilde b_{1} \left(1+ c' {y^+}^2 +d'{y^+}^3 + e'{y^+}^4 \right).
\label{extendedTS}
\end{equation}
Figure~\ref{logudivU} (right) shows that this expansion can provide a good fit to the data for $y^+<25$ ($Re_\tau > 550$) but it rapidly deteriorates further from the wall.

\section{Conclusions}

Given that $\overline{u^2}$ in the very near wall region scales with $\overline{{\tau'_{wx}}^2}$ rather than $\tau_w$, we need to revisit our dimensional analysis.  For $y^+<25$ it was shown that 
\begin{equation}
\overline{{u^2}^+} = \overline{b_1^2} (Re_\tau) f(y^+).
\label{fg}
\end{equation}
Equation~\ref{fg} may be contrasted with the much less specific  but widely assumed relationship given in equation~\ref{dim_analysis}. The analysis was performed for channel flow data in the range $550 \le Re_\tau \le 5200$, but the extension to boundary layers at higher Reynolds numbers seems robust. The results demonstrate how the process of modulation and superimposition by the large-scale motions governs the scaling of the turbulence intensity in the inner region of wall-bounded flows.

\subsection*{Acknowledgments}
This work was supported by ONR under Grant N00014-17-1-2309 (Program Manager Peter Chang). We thank Myoungkyu Lee and Robert Moser for sharing their data (available at http://turbulence.ices.utexas.edu).

\bibliographystyle{jfm}
\bibliography{BigBib_master_12_15_20}

\end{document}